\documentclass[12pt]{article}
\usepackage{a4wide,epsfig,psfrag,amsmath,amssymb,cite,scalefnt}
\usepackage{color}

\newcommand{\gsim}{\;\rlap{\lower 3.5 pt \hbox{$\mathchar \sim$}} \raise 1pt
 \hbox {$>$}\;}
\newcommand{\lsim}{\;\rlap{\lower 3.5 pt \hbox{$\mathchar \sim$}} \raise 1pt
 \hbox {$<$}\;}

\newcommand{\msusy}{m_{\rm susy}}

\newcommand{\xts}{x_{ts}}
\newcommand{\xmus}{\frac{\mu_{\rm susy}}{\msusy}}
\newcommand{\lmMes}{l_{\varepsilon}}
\newcommand{\lmMsusy}{l_{\rm susy}}
\newcommand{\lmMt}{l_{\rm t}}

\begin{document}

\title{\vskip-3cm{\baselineskip14pt
    \begin{flushleft}
      \normalsize SFB/CPP-10-111\\
      \normalsize TTP10-47
  \end{flushleft}}
  \vskip1.5cm
  Towards Higgs boson production in gluon fusion to NNLO in the MSSM
}

\author{
  Alexey Pak,
  Matthias Steinhauser,
  Nikolai Zerf
  \\[1em]
  {\small\it Institut f{\"u}r Theoretische Teilchenphysik}\\
  {\small\it Karlsruhe Institute of Technology (KIT)}\\
  {\small\it 76128 Karlsruhe, Germany}
}

\date{}

\maketitle

\thispagestyle{empty}

\begin{abstract}
We consider the Higgs boson production in the gluon-fusion channel
to next-to-next-to-leading order within the Minimal Supersymmetric Standard Model.
In particular, we present analytical results for the matching coefficient of
the effective theory and study its influence on the total production cross
section in the limit where the masses of all MSSM particles coincide.
For supersymmetric masses below 500~GeV it is possible to find
parameters leading to a significant enhancement of the Standard Model cross
section, the $K$-factors, however, change only marginally.
\medskip


\end{abstract}

\thispagestyle{empty}


\newpage


\section{Introduction}

The hunt for the Higgs boson is performed with great enthusiasm both
at the Fermilab Tevatron and at the CERN Large Hadron Collider (LHC).
The Higgs boson constitutes the last missing piece in the Standard Model
(SM) of particle physics and could explain the generation of mass terms
in the Lagrange density. The latter is also true in many extensions
of the SM, such as supersymmetry.

In the recent years many higher order corrections to production and
decay of the Higgs boson have been computed. 
The leading order (LO) and the next-to-leading order (NLO)
  corrections to the production cross section in gluon fusion
  have been considered in Refs.
  ~\cite{Georgi:1977gs,Djouadi:1991tka,Dawson:1990zj,Spira:1995rr}.
The most recent
computations within the SM that appeared in the end of
2009~\cite{Harlander:2009mq,Pak:2009dg,Harlander:2010my}
describe the Higgs boson production in gluon fusion at the
next-to-next-to-leading order (NNLO) taking the finite top
quark mass into account. An important outcome of these
papers is the justification to use the effective theory (built for
the infinite top quark mass) for the evaluation of the cross
section as it was done in
Refs.~\cite{Harlander:2002wh,Anastasiou:2002yz,Ravindran:2003um}.
As long as one restricts the analysis to the top quark/squark sector
it is suggestive that this conclusion also holds for the Minimal
Supersymmetric Standard Model (MSSM) since the corresponding
calculation would have a similar structure.
At NLO this has been demonstrated in
Refs.~\cite{Harlander:2004tp,Anastasiou:2008rm}. In this paper we
assume that it also holds at NNLO in situation when all
supersymmetric particles are heavier than the top quark.
Note that we explicitly exclude the bottom quark/squark sector from our
consideration since in that case the effective-theory approach as presented
in this paper is not applicable.\footnote{See, e.g., 
Refs.~\cite{Degrassi:2010eu,Harlander:2010wr} where the bottom
quark/squark sector is discussed at NLO in detail.} 
Thus, our results only apply to
small and moderate values of $\tan\beta$, the ratio
of the vacuum expectation values of the two Higgs doublets,
where the contribution of the bottom quark/squark sector is numerically negligible.

The supersymmetric NLO correction to the gluon fusion process has
been considered in various papers. In
Refs.~\cite{Harlander:2003bb,Harlander:2004tp} the Higgs boson production cross
section and the decay rate have been computed in the effective theory
framework and thus apply to Higgs boson masses less than about
200~GeV. The results have been confirmed by a (numerical) calculation
in the full theory~\cite{Anastasiou:2008rm} (see also 
Refs.~\cite{Degrassi:2008zj,Muhlleitner:2010nm}). It should be mentioned
that the squark contribution has been considered separately in
Refs.~\cite{Bonciani:2007ex,Muhlleitner:2006wx}. Furthermore, an
estimate of the NNLO supersymmetric corrections was presented in
Ref.~\cite{Harlander:2003kf} (see also Ref.~\cite{Harlander:2004tp}) 
assuming that the NNLO corrections to the
matching coefficient in the MSSM and the SM are numerically of the same order.

In this paper we consider the strong interaction part of the MSSM and evaluate
the total production cross section of an intermediate mass Higgs boson
in the process of gluon fusion. As motivated above, we employ the effective
theory obtained in the formal limit where all supersymmetric particles
and the top quark are infinitely heavy, which leads to the Lagrange density
\begin{eqnarray}
  {\cal L}_{Y,\rm eff} &=& -\frac{\phi^0}{v^0} C_1^0 {\cal O}_1^0 
  + {\cal L}_{QCD}^{(5)}
  \,,
  \label{eq::leff}
\end{eqnarray}
with
\begin{eqnarray}
  {\cal O}_1^0 &=& \frac{1}{4} G_{\mu\nu}^0 G^{0,\mu\nu}\,.
\end{eqnarray}
$C_1^0$ is the coefficient function containing the remnant
contributions of the heavy particles and ${\cal O}_1^0$ is the effective
operator. ${\cal L}_{QCD}^{(5)}$ is the QCD Lagrange density
with five active flavours.
In our case $\phi^0$ denotes the light CP-even Higgs
boson of the MSSM. $G_{\mu\nu}^0$ is the gluonic field strength
tensor of the standard five-flavour QCD and the superscript ``0''
indicates bare quantities.  Note that
the renormalization of $\phi^0/v^0$ is of higher order in the
electromagnetic coupling constant.

The renormalization of ${\cal O}_1^0$ is discussed
in detail in Refs.~\cite{Spiridonov:1984br,Chetyrkin:1997un}.
For completeness we reproduce here the renormalization constant
which in the $\overline{\rm MS}$ scheme is given by
\begin{eqnarray}
  {\cal O}_1 = Z_{11}\,{\cal O}_1^0\,,\qquad
  Z_{11} = \left(1 - \frac{\pi}{\alpha_s}\frac{\beta^{(5)}}{\epsilon}\right)^{-1}\,,
  \label{eq::c1ren}
\end{eqnarray}
where $d=4-2\epsilon$ is the space-time dimension and
$\beta^{(5)}=\beta^{(5)}(\alpha_s)$ is the $\beta$ function of the standard
five-flavour QCD with $\beta^{(5)} =
-(\alpha_s^{(5)}/\pi)^2\beta_0^{(5)}+\ldots$ 
and $\beta_0^{(n_l)}=11/4-n_l/6$.
Since $C_1^0 {\cal O}_1^0 = C_1 {\cal O}_1$, the coefficient
function is renormalized with $1/Z_{11}$.

Following Ref.~\cite{Chetyrkin:1996ke}, we slightly
redefine the coefficient function and the operator in order to have
objects which are separately renormalization group invariant.
Introducing
\begin{eqnarray}
  {\cal O}_g &=& B^{(5)} {\cal O}_1 \,,\nonumber\\
  C_g &=& \frac{1}{B^{(5)}} C_1\,,
  \label{eq::B5}
\end{eqnarray}
with
\begin{eqnarray}
  B^{(5)} &=& -\frac{\pi^2 \beta^{(5)}}{\beta_0^{(5)}\alpha_s^{(5)}}
  \,,
\end{eqnarray}
we may choose independently the ``soft'' and the ``hard''
renormalization scales. Our following analysis of the Higgs
production cross section takes advantage of
those properties of ${\cal O}_g$ and $C_g$.
Note that the LO piece of $C_g$ is not proportional to $\alpha_s$.

In this paper we consider the strongly interacting part of the MSSM and assume
for simplicity the degenerate supersymmetric mass spectrum. The evaluation of
the coefficient functions is performed within Dimensional Reduction
({\tt DRED})~\cite{Siegel:1979wq} where we follow
Refs.~\cite{Harlander:2009mn,Kant:2010tf} for the practical 
implementation.

The remainder of the paper is organized as follows: in the following
section we briefly revisit the SM and in particular discuss the effect of
$B^{(5)}$ introduced above. We furthermore evaluate the SM coefficient
function within {\tt DRED} which is a useful preparation for the SUSY
QCD calculation.
Next, in Section~\ref{sec::mssm} we discuss
the three-loop calculation of the MSSM coefficient function
which is used in Section~\ref{sec::xsec} to compute the cross
section to NNLO. Section~\ref{sec::concl} contains the summary and conclusions.


\section{\label{sec::sm}Revisiting the SM}

As mentioned in the Introduction, the gluon fusion process has been discussed
extensively in the literature. The purpose of this section is twofold: on the
one hand we want to discuss the effect of $B^{(5)}$ introduced in
Eq.~(\ref{eq::B5}) on the numerical predictions, on the other hand we
repeat the SM calculation in the framework of {\tt DRED} as a preparation for
the calculation in the supersymmetric theory.

\subsection{Separation of hard and soft scales}

The SM coefficient function expressed in terms of the
on-shell top quark mass and the five-flavour strong coupling
reads\footnote{Here and in the following we suppress the $\mu$-dependence
  of $\alpha_s$. If not stated otherwise we have
  $\alpha_s\equiv\alpha_s(\mu)$. The same is true for the $\overline{\rm DR}$
  mass parameters $m_t$ and $\msusy$ introduced in
  Section~\ref{sec::mssm}.}\cite{Chetyrkin:1997un,Steinhauser:2002rq} (see
also Ref.\cite{Kramer:1996iq}):
\begin{eqnarray}
  C_1^{\rm SM} &=& -\frac{1}{3} \frac{\alpha_s^{(5)}}{\pi} \Bigg\{
    1 + \frac{\alpha_s^{(5)}}{\pi} \frac{11}{4}
    \nonumber \\
    &&\mbox{}+ \left(\frac{\alpha_s^{(5)}}{\pi}\right)^2
    \left[\frac{2777}{288} + \frac{19}{16}\ln\frac{\mu^2}{M_t^2}
      + n_l \left(- \frac{67}{96} + \frac{1}{3}\ln\frac{\mu^2}{M_t^2}\right)
      \right]
  \Bigg\}\,,
  \label{eq::C1SM}
\end{eqnarray}
where $n_l=5$ is the number of massless quarks.

\begin{figure}[t]
  \centering
  \begin{tabular}{c}
    \includegraphics[width=0.75\linewidth]{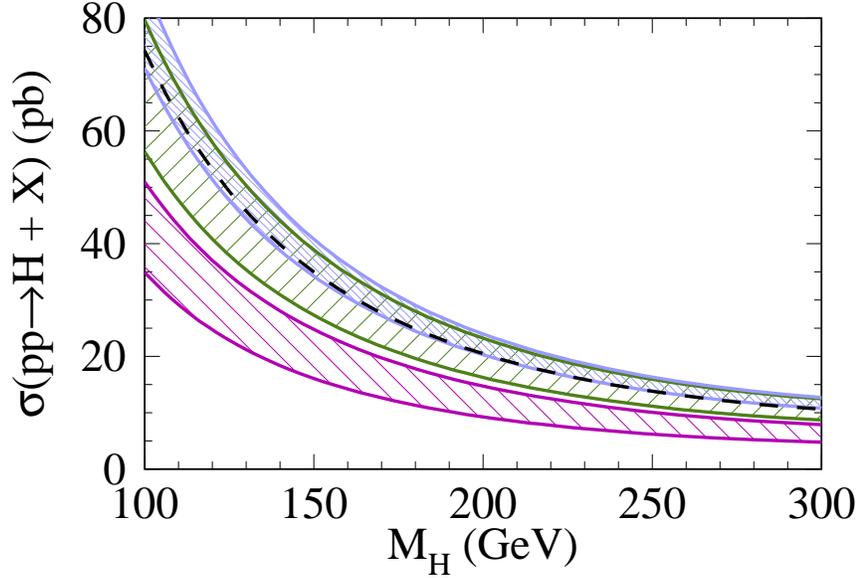}
    \\ (a) \\[-1em]
    \includegraphics[width=0.75\linewidth]{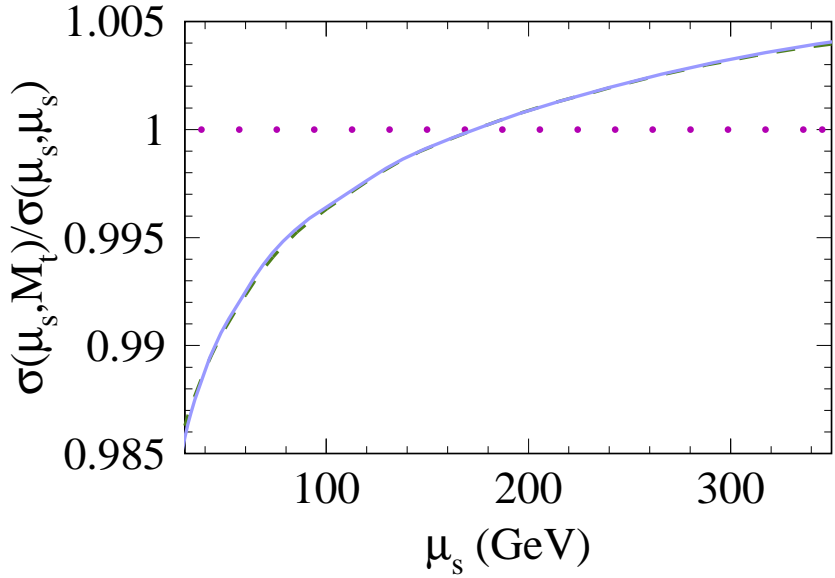}
    \\ (b)
  \end{tabular}
  \caption[]{\label{fig::xsecsm}
    (a) The SM cross section as a function of $M_H$.
    The LO, NLO, and NNLO results are represented by the lower, middle
    (coarse filling) and the upper bands, respectively. Band width
    is due to simultaneous variation of renormalization and factorization
    scales. The dashed line corresponds to the NNLO prediction using ABKM09.
    (b) Ratio $R=\sigma(\mu_s,M_t)/\sigma(\mu_s,\mu_s)$ for $M_H=120$~GeV
    as a function of $\mu_s$. Dotted, dashed (barely visible) and solid lines
    correspond to LO, NLO and NNLO, respectively.
    }
\end{figure}

In Fig.~\ref{fig::xsecsm}(a) we show the production cross section
$\sigma(pp\to H+X)$ for the LHC with the center-of-mass energy of 14~TeV
at LO, NLO and NNLO for $M_t=173.3$~GeV~\cite{:1900yx},
computed with MSTW08~\cite{Martin:2009iq} parton distribution function (PDF)
set. 

The chosen PDFs fix the values of $\alpha_s^{(5)}(M_Z)$ to LO, NLO and NNLO
accuracy used as the starting point in order to obtain the strong
coupling for the other choices of the renormalization scale.
The bands are obtained by varying the renormalization and factorization
scales $\mu_r = \mu_f = \mu$ between $M_H/4$ and
$M_H$~\cite{Anastasiou:2008tj,Grazzini:2010zc}\footnote{For the other 
opinions concerning the scale variation see Ref.~\cite{Baglio:2010um}.}
with $\mu=M_H/2$ as central value.
For comparison we present the NNLO result computed with ABKM09
PDFs~\cite{Alekhin:2009ni} (the dashed line). It is remarkable that the
difference in the central values for the different PDFs is comparable to the
uncertainty due to the scale variation.

The three-loop term of Eq.~(\ref{eq::C1SM}) contains the
logarithm $\ln(\mu^2/M_t^2)$ which can potentially lead to numerically
big effects if the scale $\mu$ assumes very small or very large values.
With the help of $C_g$ and ${\cal O}_g$ introduced in Eq.~(\ref{eq::B5})
we may introduce separate scales $\mu_h$ and $\mu_s$
for the hard and the soft parts of the cross section
$\sigma=\sigma(\mu_s,\mu_h)$. The dangerous logarithms can
then be avoided by choosing $\mu_h \sim M_t$ which is the natural
scale for the coefficient function, and varying $\mu_s$ around $M_H/2$
as done above. Note that for $\mu_s=\mu_h$ the quantity $B^{(5)}$ drops
out of the analysis.

In Fig.~\ref{fig::xsecsm}(b) we plot the ratio
$R=\sigma(\mu_s,M_t)/\sigma(\mu_s,\mu_s)$ for $M_H=120$~GeV.
In the numerator the hard scale is kept fixed while in
the denominator we use the common
approach~\cite{Harlander:2002wh,Anastasiou:2002yz,Ravindran:2003um}
and identify the scales as in Fig.~\ref{fig::xsecsm}(a).
At LO $R=1$ by construction. The NLO corrections (dashed curve which is hardly
visible) are quite small and amount to less than $\pm 2\%$ which is only 
slightly modified after inclusion of the NNLO results (solid line).
From these considerations we conclude that the resummation of
$\ln(\mu^2/M_t^2)$ terms has only a small numerical effect.
On the other hand, the framework of
Eq.~(\ref{eq::B5}) is quite useful in the context of supersymmetric
corrections discussed below since it allows to fix the hard scale $\mu_h$
and vary $\mu_s$.

\subsection{\label{sub::SMDRED}$C_1^{\rm SM}$ in {\tt DRED}}

Whereas {\tt DRED} is the preferable framework for the two- and three-loop
calculations in a supersymmetric theory, in the SM the application of {\tt DRED}
introduces several so-called evanescent couplings which can make practical
calculations quite tedious. This has been discussed extensively in the 
literature (see, e.g.,
Refs.~\cite{Jack:1993ws,Jack:1994bn,Harlander:2006rj,Harlander:2006xq,Jack:2007ni}).  
Nevertheless, for our purpose it is useful to evaluate the QCD corrections to
the matching coefficient using {\tt DRED} since there are several new features
which also apply to the MSSM.

We implement {\tt DRED} with the help of the so-called $\varepsilon$
scalars. They have mass dimension $(4-d)$ and complement the
$d$-dimensional gluon field so that the gauge fields and the fermions
have exactly four degrees of freedom.
In a practical calculation the $\varepsilon$ scalars induce new Feynman
rules (see, e.g., Ref.~\cite{Kant:2009zza})
which (in non-supersymmetric theories) involve new couplings
that have to be renormalized independently from the gauge coupling.

In our matching calculation it is convenient to define $C_1^{\rm SM}$ 
in such a way that the effective theory is
built from the five-flavour QCD fields regularized with {\tt DREG}. As a consequence,
we have to integrate the $\varepsilon$ scalar field out together with the top
quark in order to guarantee the transition from {\tt DRED} to {\tt DREG} in
the matching procedure.
This means that formally $M_\varepsilon\not=0$ and even 
$M_\varepsilon\gg M_H$ since the latter is identical to zero for the matching
calculation. In the end, we take the limit $M_\varepsilon\to 0$ in agreement with
our renormalization condition (see below).

We evaluate the matching coefficient in the limit
$M_t\gg M_\varepsilon\not=0$ and apply the rules of asymptotic expansion together
with a naive expansion in the external momenta in order to project out the
Lorentz structure of the Higgs-gluon-gluon vertex (see the following section for
details).
Consequently, there are (sub-)diagrams which contain only 
$\varepsilon$ scalars as massive particles.
Due to the dimension of the underlying integrals they lead to 
terms proportional to $M_\varepsilon^{-2}$ and $M_\varepsilon^{-4}$
in the sum of the bare diagrams. Thus it is
necessary to keep positive powers in $M_\varepsilon$ in counterterms
entering the calculation.

The top quark mass and the strong coupling need to be renormalized
at two-loop order. Furthermore, one has to make sure
that the external gluon fields of the full and the effective theory
Green functions are identical.
There are two operators involved in the renormalization procedure which
deserve additional attention: the $\varepsilon$-scalar mass and the
coupling of the Higgs boson to $\varepsilon$ scalars. The corresponding
part of the Lagrange density is
\begin{eqnarray}
  {\cal L}_\varepsilon &=& 
  -\frac{1}{2}\left(M_\varepsilon^0\right)^2 \varepsilon^{0,a}_\sigma\varepsilon^{0,a}_\sigma
  - \frac{\phi^0}{v^0} \left(\Lambda_\varepsilon^0\right)^2
  \varepsilon^{0,a}_\sigma\varepsilon^{0,a}_\sigma 
  + \ldots
  \,.
  \label{eq::Lep}
\end{eqnarray}
As mentioned above, we keep $M_\varepsilon \ne 0$ for the
matching procedure, however, in the final expression for $C_1^{\rm SM}$ we
require that the on-shell $M_\varepsilon$ vanishes.
This leads to the well-known counterterm 
(see, e.g., Refs.~\cite{Marquard:2007uj,Kant:2010tf})
which we need to one-loop accuracy. For our purposes we have to
keep the terms proportional to $M_\varepsilon^2$ and restore
the dependence on all evanescent couplings.

The second term in Eq.~(\ref{eq::Lep}) is
not present at tree-level, however, quantum corrections induce such a
structure~\cite{Anastasiou:2008rm}. We require that the
renormalized parameter $\Lambda_\varepsilon$ vanishes since we
want to avoid a non-zero coupling between the Higgs boson and the 
unphysical $\varepsilon$ scalars. As a consequence of this condition,
we have to introduce 
a non-zero bare coupling of the Higgs boson to $\varepsilon$ scalars. 
Furthermore, in the counterterm contribution to
$\Lambda_\varepsilon$ we only have to keep terms not 
proportional to $\Lambda_\varepsilon$.
We refrain from listing the explicit results which become
quite lengthy due to the evanescent couplings.

It is an important check of our calculation that the
negative powers in $M_\varepsilon$ cancel after
the renormalization of $\Lambda_\varepsilon$ and $M_\varepsilon$.

Our final result for the coefficient function reads:
\begin{align}
 C_1^{\overline{\text{DR}}}=& -\frac{\alpha_s^{\overline{\text{DR}}}}{3\pi}\Bigg\{
   1 
  +\frac{\alpha_s^{\overline{\text{DR}}}}{\pi} \bigg(
          \frac{5}{2} 
         -\frac{1}{6}l_t
   \bigg) 
  +\frac{\alpha_s^{\overline{\text{DR}}}\alpha_e}{\pi^2}\bigg(
       \frac{7}{36} 
      +n_l\frac{1}{12}
   \bigg)\nonumber\\
  & +\left(\frac{\alpha_s^{\overline{\text{DR}}}}{\pi}\right)^2\Bigg[
      \frac{2065}{288} 
      -\frac{5}{48} l_t 
      +\frac{1}{36}l_t^2 
      -n_l\bigg(
      \frac{67}{96}
     -\frac{1}{3}l_t
   \bigg)
   \Bigg]
  \Bigg\}
  \,,
  \label{eq::C1DR_QCD}
\end{align}
where $\lmMt=\ln[\mu^2/(m_t(\mu^2))^2]$, $\alpha_s^{\overline{\rm DR}}$
is the strong coupling with six
active flavours in the $\overline{\rm DR}$ scheme and $\alpha_e$ is
the evanescent coupling originating from the $\varepsilon$
scalar-quark vertex. Note that the evanescent coupling stemming from the
four-$\varepsilon$-vertex cancels in Eq.~(\ref{eq::C1DR_QCD}).

The transition to the five-flavour QCD is achieved with the help of
\begin{align}
\alpha_s^{\overline{\rm DR}}=&\,\, \alpha_s^{(5)}\zeta_{\alpha_s}^{-1}
   \,,
  \nonumber\\
  \zeta_{\alpha_s}^{-1} =&\,\,
  1 
  + \frac{\alpha_s^{(5)}}{\pi}
  \left[\frac{1}{4}+ \frac{1}{6}\lmMt + \epsilon\left(\frac{1}{4}\lmMes+\frac{1}{12}\lmMt^2 + \frac{1}{12}\zeta_2\right)\right]\nonumber\\
  &+ \left(\frac{\alpha_s^{(5)}}{\pi}\right)^2
  \left(\frac{11}{9} + \frac{13}{24}\lmMt + \frac{1}{36}\lmMt^2\right) 
  -\frac{\alpha_s^{(5)}\alpha_e}{\pi^2}\left(\frac{7}{36}+n_l\frac{1}{12}\right)
  \,.
  \label{eq::zetag_QCD}
\end{align}
From Eqs.~(\ref{eq::C1DR_QCD}) and~(\ref{eq::zetag_QCD})
we recover $C_1^{\rm SM}$ of Eq.~(\ref{eq::C1SM})\footnote{Note that the
  functional form of Eq.~(\ref{eq::C1SM}) is independent from the quark
  mass definition.} which
is a welcome cross check on all individual steps of our calculation,
in particular on the counterterms originating from
  Eq.~(\ref{eq::Lep}). In the next section the same framework is applied
  to the MSSM. 


\section{\label{sec::mssm}Coefficient function in the MSSM to three loops}

In this section we describe the evaluation of the coefficient function $C_1$
of Eq.~(\ref{eq::leff}) within the strong sector of the MSSM. Note that the operator
${\cal O}_1$ contains the effective interaction of two, three or four
gluons with the Higgs boson and we are free to choose any
corresponding Green functions in order to extract $C_1$. We
decided to consider the two-gluon-Higgs boson amplitude since
it involves fewer diagrams. We need to keep the two gluon momenta
different until the Lorentz structure of $O_1$ responsible for
the two-gluon coupling is projected out.

\begin{figure}[t]
  \centering
  \includegraphics[width=\linewidth]{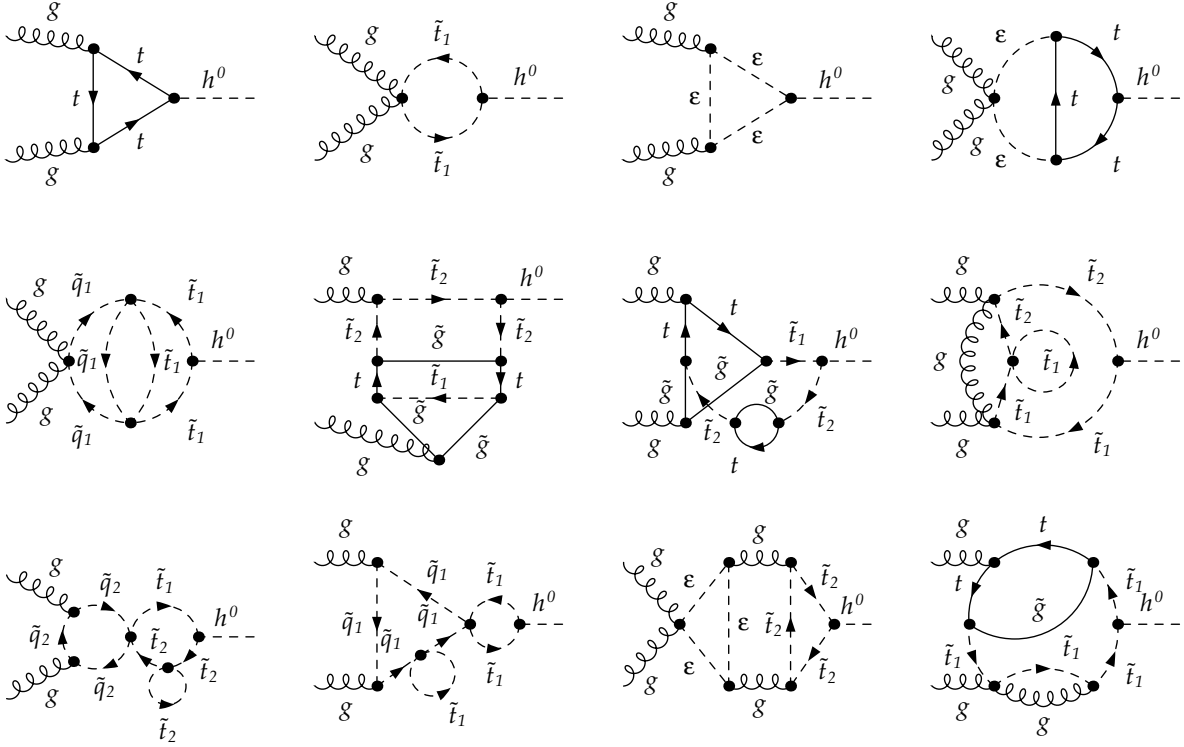}
  \caption[]{\label{fig::diags}Sample diagrams contributing to 
    $C_1$ at one, two and three loops.
    The symbols $t$, $\tilde{t}_i$, $g$, $\tilde{g}$, $h^0$ and
    $\varepsilon$ denote top quarks, top squarks, gluons, gluinos,
    Higgs bosons and $\varepsilon$ scalars, respectively.
    }
\end{figure}

At one-, two- and three-loop order 10, 671 and 49632 Feynman
diagrams have to be considered. Some typical diagrams are shown
in Fig.~\ref{fig::diags}.
In addition to the particles of
supersymmetric QCD we also have the $\varepsilon$ scalar.
Higgs boson couplings to the light quarks and the corresponding
squarks are omitted. Furthermore, we neglect the contribution of the
Higgs boson-squark coupling suppressed by $M_Z^2/m_t^2$ (for
the Feynman rules see, e.g., Appendix~A of Ref.~\cite{Harlander:2004tp}).

Our calculation is based on a chain of programs designed to minimize
the manual input. First, we generate all the Feynman diagrams with
{\tt QGRAF}~\cite{Nogueira:1991ex}, then transform the
output to {\tt FORM}~\cite{Vermaseren:2000nd} notation
with the help of {\tt q2e} and apply asymptotic expansion
(see, e.g., Ref.~\cite{Smirnov:2002pj})
using {\tt exp}~\cite{Harlander:1997zb,Seidensticker:1999bb}.

Due to the gauge invariance the Higgs-gluon-gluon coupling has the form
\begin{eqnarray}
  {\cal A}^{\mu\nu,ab} &=& A \,\, \delta^{ab}
  \left(g^{\mu\nu}q_1\cdot q_2 - q_1^\nu q_2^\mu\right)\,,
\end{eqnarray}
where $A$ is a scalar function depending on $M_H^2=2q_1\cdot q_2$, $M_t^2$,
$\mu^2$, and the mass parameters of SUSY QCD. $q_1$ and $q_2$ are the
outgoing momenta of the gluons with polarization vectors $\varepsilon^\mu(q_1)$
and $\varepsilon^\nu(q_2)$ and colour indices $a$ and $b$.
We begin by convoluting the vertex Feynman diagrams with the projector
\begin{eqnarray}
  P^{ab}_{\mu\nu}(q_1,q_2) &=& 
  \delta^{ab}
  \frac{q_1\cdot q_2 g_{\mu \nu}
    -q_{1\mu} q_{2\nu}
    -q_{1\nu} q_{2\mu}}
       {8(d-2)(q_1\cdot q_2)^2}
  \,,
  \label{eq::proj}
\end{eqnarray}
so that only scalar integrals remain.
After the Taylor expansion to the linear order in $q_1$ and $q_2$
the factor $(q_1\cdot q_2)^2$ in the denominator of Eq.~(\ref{eq::proj})
cancels and both momenta can be set to zero.
Finally, we evaluate vacuum integrals which (after the asymptotic
expansion) only contain a single scale. This is done using the package
{\tt MATAD}~\cite{Steinhauser:2000ry}.

Due to the transverse structure of the Higgs-gluon-gluon amplitude 
${\cal A}^{\mu\nu,ab}$ 
an alternative projector to Eq.~(\ref{eq::proj}) can be applied
to obtain $C_1$:
\begin{eqnarray}
  \tilde{P}^{ab}_{\mu\nu}(q_1,q_2) &=& 
  \delta^{ab}
  \frac{q_1\cdot q_2 g_{\mu \nu}
    -(d-1)q_{1\mu} q_{2\nu}
    -q_{1\nu} q_{2\mu}}
       {8(d-2)(q_1\cdot q_2)^2}
  \,.
  \label{eq::proj2}  
\end{eqnarray}
We have checked that in both cases we obtain the same result.

One may consider different hierarchies between the mass scales
involved in this calculation.
As the first choice, we identify all supersymmetric masses with
$\msusy$ and thus have two expansion parameters, $x_{ts} = m_t/\msusy$
and $x_{\varepsilon t}=M_{\varepsilon}/m_t$.
At three-loop order we have computed terms through
${\cal O}\left(x_{ts}^6\right)$ and ${\cal O}\left(x_{\varepsilon t}^0\right)$.
As a cross-check, we performed a similar calculation where the top quark
mass and $\msusy$ were identified from the very beginning. 
Comparing the latter result to the former expansion in $\xts$ evaluated at
$\xts = 1$ we find agreement to within a few per cent, which
serves as a convergence test of our results.

We choose to renormalize all parameters in the 
$\overline{\rm DR}$ scheme~\cite{Siegel:1979wq}. In particular, we need
the counterterms for
$\alpha_s$, the top quark, and the top squark masses
to two-loop, and the gluino and other squark mass
counterterms to one-loop order. Furthermore, the
counterterm for the mixing angle $\theta_t$ between the top squarks
is also required to the two-loop order. Although $\theta_t$ is zero in our
approximation, its renormalization in the non-degenerate one- and
two-loop expressions leads to non-zero contributions at three-loop
level. The corresponding counterterm can either be obtained from
the exact one- and two-loop result for
$C_1$~\cite{Harlander:2004tp}, or (as done in this paper) via expansion 
in the mass difference between the two top squarks. Note that the
same effect is also present in the
three-loop calculation of the Higgs boson mass within the MSSM, as
discussed in Ref.~\cite{Harlander:2008ju,Kant:2010tf}.
All necessary counterterms can be found in the Appendix of
Ref.~\cite{Kant:2010tf} except for the decoupling constant relating
the gluon field in the SUSY-QCD to that in the five-flavour QCD, given by
\begin{align}
\zeta_3^0=&  1 
   +\frac{\alpha_s^{(\text{SQCD})}}{\pi}
 \Bigg\{
     \frac{1}{\epsilon}
     \bigg(
         \frac{3}{4} 
       + \frac{n_l}{12}
     \bigg)
   + \frac{1}{4} 
   + \frac{7}{12}l_{\text{susy}} 
   + \frac{1}{6}l_t 
   + n_l\frac{1}{12}l_{\text{susy}}\nonumber\\
 & + \epsilon 
 \Bigg[
 \frac{1}{4}l_{\varepsilon} 
  + \frac{7}{24}l_{\text{susy}}^2 
  + \frac{1}{12}l_t^2 
  + \frac{3 \zeta_2}{8}
  + n_l 
 \bigg(
  \frac{1}{24}l_{\text{susy}}^2 
    + \frac{\zeta_2}{24}
  \bigg) 
  \Bigg]
  \Bigg\}\nonumber\\
 & +\left(\frac{\alpha_s^{(\text{SQCD})}}{\pi}\right)^2 
    \Bigg\{
 \frac{1}{\epsilon^2}
     \bigg(
   - \frac{69}{64} 
   + n_l\frac{13}{192} 
  + n_l^2\frac{1}{48}
  \bigg)\nonumber\\
  & +\frac{1}{\epsilon}
    \Bigg[
     \frac{421}{384} 
    - \frac{49}{96} l_{\text{susy}} 
    - \frac{7}{48} l_t
    + n_l
  \bigg(
     - \frac{25}{384} 
     + \frac{7}{96} l_{\text{susy}}
     + \frac{1}{24}l_t
   \bigg)
    + n_l^2\frac{1}{48}l_{\text{susy}}
 \Bigg]\nonumber\\
 & + \frac{1535}{2304} 
  - \frac{7}{32} l_{\varepsilon} 
   + \frac{53}{192} l_{\text{susy}} 
   + \frac{7}{96} l_{\text{susy}}^2 
   + \frac{29}{96} l_t 
   + \frac{1}{48}l_t^2
   - \frac{21 \zeta_2}{64}\nonumber\\
 & + n_l 
 \bigg(
    \frac{97}{2304} 
   + \frac{1}{16}l_{\varepsilon} 
   - \frac{67}{576} l_{\text{susy}} 
   + \frac{1}{12}l_{\text{susy}}^2 
   + \frac{1}{48}l_t^2 
   + \frac{11 \zeta_2}{192}
 \bigg)\nonumber\\ 
 & + n_l^2 
 \bigg(
 \frac{1}{96}l_{\text{susy}}^2 
   + \frac{\zeta_2}{96}
 \bigg) 
   + x_{ts}^2
 \bigg(
   -\frac{1}{432} 
   - \frac{13}{72} l_{\text{susy}} 
   + \frac{5}{72} l_t
 \bigg)  
 \Bigg\} 
 + {\cal O}\left(\xts^4\right)\,,
\label{eq::zeta30}
\end{align}
where $\lmMsusy=\ln[\mu^2/(\msusy(\mu^2))^2]$ and 
$\alpha_s^{\rm (SQCD)}$ denotes the $\overline{\mbox{DR}}$ coupling
where all particles of the theory contribute to the running.
In Eq.~(\ref{eq::zeta30}) $n_l=5$ counts the number of massless quarks and
at the same time the number of squark flavours different from top squarks.
Similarly to the SM case in Section~\ref{sub::SMDRED}, we keep $M_\varepsilon \ne 0$
in the intermediate steps
in order to match to the effective theory of Eq.~(\ref{eq::leff}).
Thus, $M_\varepsilon$ and the 
coupling $\Lambda_\varepsilon$ (cf. Eq.~(\ref{eq::Lep})) have to be renormalized
with the help of the counterterms
\begin{eqnarray}
  \left(M^0_{\varepsilon}\right)^2 &=&
  M_{\varepsilon}^2 +
  \frac{\alpha_s^{\rm(SQCD)}}{\pi}
  \Bigg\{m_t^2\left(\frac{1}{2\epsilon}+\frac{1}{2}+\frac{1}{2}\lmMt\right)
    + M_{\varepsilon}^2\Bigg[
      \frac{8 - n_l}{4\epsilon} 
    + \frac{9}{2} 
    + 3\lmMes 
    - \frac{3}{4}\lmMsusy 
    - \frac{1}{4}\lmMt 
    \nonumber\\&&\mbox{}
    + \left(-\frac{1}{2} - \frac{1}{4}\lmMes\right)n_l
    \Bigg]
    + m_{\rm susy}^2
    \left[1 + \lmMsusy + \frac{2 - nl}{2\epsilon} 
    + \left(-\frac{1}{2} - \frac{1}{2}\lmMsusy\right)n_l
    \right]
    \Bigg\}
    \,,
    \nonumber\\
\end{eqnarray}
with $\lmMes=\ln[\mu^2/M_\varepsilon^2]$, and
\begin{align}
\left(\Lambda_\varepsilon^0\right)^2 - \Lambda_\varepsilon^2=&\,\,\delta \Lambda_{\varepsilon}^2
\nonumber\\
=&-2m_t^2\frac{c_\alpha}{s_{\beta}} 
                            \Bigg\{
                                      \frac{\alpha_s^{(\text{SQCD})}}{\pi}
                                      \Bigg[
                                               \frac{1}{2}l_{\text{susy}} 
                                             -  \frac{1}{2}l_t 
                                             + \epsilon 
                                               \bigg(
                                                      \frac{1}{4}l_{\text{susy}}^2 
                                                     - \frac{1}{4}l_t^2
                                               \bigg)
                                       \Bigg]\nonumber\\
                                  & + \left(\frac{\alpha_s^{(\text{SQCD})}}{\pi}\right)^2 
                                      \Bigg[
                                              \frac{1}{\epsilon}
                                               \bigg(
                                                      -l_{\text{susy}}
                                                      +l_t 
                                                      +n_l
                                                       \Big(
                                                              \frac{1}{8}l_{\text{susy}} 
                                                            - \frac{1}{8}l_t
                                                       \Big) 
                                               \bigg)\nonumber\\
                                       &     + \frac{1}{3} 
                                             + \frac{23}{12} l_{\text{susy}} 
                                             - \frac{1}{8}l_{\text{susy}}^2 
                                             - \frac{23}{12} l_t 
                                             - \frac{1}{12}l_{\text{susy}} l_t 
                                             + \frac{5}{24} l_t^2 
                                             + n_l
                                               \bigg(
                                                       \frac{1}{16}l_{\text{susy}}^2 
                                                     - \frac{1}{16}l_t^2
                                               \bigg)\nonumber\\ 
                                      &
                                             + x_{ts} x_{\mu t}
                                               \bigg(
                                                      t_{\alpha}
                                                     +\frac{1}{t_{\beta}}
                                               \bigg)
                                               \bigg(
                                                       \frac{11}{12} 
                                                     + \frac{1}{6}l_{\text{susy}} 
                                                     + \frac{1}{6}l_t
                                               \bigg)
                                               + x_{ts}^2
                                               \bigg(
                                                       -\frac{359}{216} 
                                                       +\frac{19}{72} l_{\text{susy}} 
                                                       -\frac{67}{72} l_t 
                                               \bigg) 
                                      \Bigg]
                            \Bigg\}
                            \nonumber\\&
                            + {\cal O}(\xts^4)
                            + {\cal O}(\xts^3 x_{\mu t})
                            \,,
\label{eq::Lambda_ep_SQCD}
\end{align}
where $c_\alpha=\cos\alpha$, $s_\beta=\sin\beta$,
$t_\alpha=\tan\alpha$, $t_\beta=\tan\beta$ and $x_{\mu t}=\mu_{\text{susy}}/m_t$.
Here $\alpha$ is the
mixing angle between the weak and the mass eigenstates of the neutral
scalar Higgs bosons, $\tan\beta$ is the ratio of the vacuum expectation values
of the two Higgs doublets, and $\mu_{\rm susy}$ is the Higgs-Higgsino
bilinear coupling from the super potential.
Note that the counterterm for $\Lambda_\varepsilon$ 
and the three-loop vertex diagrams have been evaluated
assuming the same hierarchy of the top quark and the top squark masses.
Let us mention that the counterterm in
Eq.~(\ref{eq::Lambda_ep_SQCD}) is finite at one-loop order and only
contains a single pole at two loops whereas the corresponding
expression in QCD has a single and a double pole, respectively.

In the first step we obtain $C_1$ in terms of $\alpha_s^{\rm (SQCD)}$.
It is, however, more convenient to express the final result in terms of
the five-flavour coupling in the $\overline{\rm MS}$ scheme
$\alpha_s^{(5)} = \alpha_s^{(5),\overline{\rm MS}}$.
The relation establishing this transition is as follows:
\begin{eqnarray}
  \alpha_s^{(5)} &=& \zeta_{\alpha_s} \alpha_s^{\rm (SQCD)}
  \,.
  \label{eq::zetaas}
\end{eqnarray}
Two-loop corrections to the decoupling constant $\zeta_{\alpha_s}$
within supersymmetric QCD have been computed in 
Refs.~\cite{Harlander:2005wm,Bauer:2008bj,Bednyakov:2007vm}. Since
$\zeta_{\alpha_s}$ in Eq.~(\ref{eq::zetaas}) combines the decoupling of the
heavy particles and the transition from $\overline{\rm DR}$ to
$\overline{\rm MS}$, we present the explicit result
in the limit of equal supersymmetric masses:
\begin{eqnarray}
  \zeta_{\alpha_s} &=& 1 
  + \frac{\alpha_s^{\rm (SQCD)}}{\pi}\left(
  -\frac{1}{4} - \frac{7}{12}\lmMsusy - \frac{1}{6}\lmMt 
  - n_l\frac{1}{12}\lmMsusy
  \right)
  + \left(\frac{\alpha_s^{\rm (SQCD)}}{\pi}\right)^2 
  \Bigg[
    \frac{77}{288} 
    \nonumber\\&&\mbox{}
    - \frac{5}{8}\lmMsusy 
    + \frac{49}{144}\lmMsusy^2
    - \frac{3}{8}\lmMt 
    + \frac{7}{36}\lmMsusy\lmMt 
    + \frac{1}{36}\lmMt^2 
    + n_l \Bigg(
    \frac{23}{144}
    - \frac{1}{72}\lmMsusy
    + \frac{7}{72}\lmMsusy^2
    \nonumber\\&&\mbox{}
    + \frac{1}{36}\lmMsusy\lmMt
    \Bigg)
    + n_l^2\frac{1}{144}\lmMsusy^2
    + \left(\frac{1}{432} + \frac{13}{72}\lmMsusy
    - \frac{5}{72}\lmMt\right)\xts^2
    \Bigg]
    + {\cal O}(\xts^4)
    \,.
\end{eqnarray}
All occurring mass parameters and the top squark mixing angle have been renormalized in the 
$\overline{\mbox{DR}}$ scheme.

Finally, we present the coefficient function $C_1$ expressed in terms
of $\overline{\mbox{DR}}$ masses in the full theory, expanded in
$\alpha_s^{(5)}$:
\begin{eqnarray}
  C_1^{\overline{\rm DR}} &=&
  -\frac{\alpha_s^{(5)}}{3\pi}\frac{c_\alpha}{s_\beta}
  \Bigg\{
  1 + \frac{\xts^2}{2} 
  + \frac{\alpha_s^{(5)}}{\pi}\Bigg[
  \frac{11}{4} - \frac{1}{3}\left(t_\alpha +
    \frac{1}{t_\beta}\right)\xmus 
  + \left(\frac{23}{12} + \frac{5}{12}\lmMsusy 
    \right.\nonumber\\&&\left.\mbox{}
    -\frac{5}{12}\lmMt\right)\xts^2
  \Bigg]
  + \left(\frac{\alpha_s^{(5)}}{\pi}\right)^2
  \Bigg[
  \frac{2777}{288} + \frac{19}{16}\lmMt + \left(-\frac{67}{96} +
    \frac{1}{3}\lmMt\right)n_l 
  \nonumber\\&&\mbox{}
  + \left(-\frac{85}{54} - \frac{85}{108}\lmMsusy -
    \frac{1}{18}\lmMt  + \left(-\frac{1}{18} +
      \frac{1}{12}\lmMsusy\right)n_l
  \right)\left(t_\alpha+\frac{1}{t_\beta}\right)\xmus
  \nonumber\\&&\mbox{}
  + \Bigg(
  \frac{30779857}{648000} - \frac{6029}{10800}\lmMsusy 
  - \frac{475}{288}\lmMsusy^2 - \frac{20407}{10800}\lmMt 
  + \frac{26}{45}\lmMsusy\lmMt - \frac{377}{1440}\lmMt^2 
  \nonumber\\&&\mbox{}
  - \frac{15971}{576}\zeta_3 
  + \left(
    - \frac{3910697}{216000} 
    + \frac{63}{4}\zeta_3
    + \frac{3307}{4800}\lmMsusy
    + \frac{131}{576}\lmMsusy^2
    + \frac{5113}{14400}\lmMt
    \right.\nonumber\\&&\left.\mbox{}
    - \frac{101}{1440}\lmMsusy\lmMt
    + \frac{3}{320}\lmMt^2
  \right) n_l
  \Bigg)
  \xts^2
  \Bigg]
  \Bigg\}
  + {\cal O}\left(\xts^4\right)
  + {\cal O}\left(\xmus\xts^2\right)
  \,.
  \label{eq::C1DR}
\end{eqnarray}
The analytic expression including ${\cal O}(\xts^6)$ and ${\cal O}(\xmus\xts^4)$ is available
on request from the authors.
It is noteworthy that in the limit $m_{\rm susy}\to\infty$ we recover exactly
the SM result of Eq.~(\ref{eq::C1SM}).


\section{\label{sec::xsec}Higgs boson production at NNLO}

In this section we discuss the numerical effect of the corrections presented
above. The effective-theory parts in the MSSM and the SM are identical,
and thus we can borrow the framework of Ref.~\cite{Pak:2009dg}. In addition,
we incorporate the factor $B^{(5)}$ (in the $\overline{\rm MS}$
scheme) as discussed in the Introduction.

On the full-theory side we consider the renormalization group invariant
combination $C_g=C_1/B^{(5)}$
where both $C_1$ and $B^{(5)}$ are parameterized with the five-flavour
strong coupling in the $\overline{\rm MS}$ scheme.
For the numerical evaluation of $C_1^{\overline{\rm DR}}$ we fix
the (hard) renormalization scale at the on-shell top quark mass value,
i.e. $\mu_h=M_t$, which means that the $\overline{\rm DR}$ mass parameters in
Eq.~(\ref{eq::C1DR}) have to be interpreted as $m_{\rm susy}(M_t)$ and
$m_t(M_t)$.
Any variation of $\mu_h$ leads to a non-degenerate supersymmetric spectrum
and thus renders Eq.~(\ref{eq::C1DR}) not applicable.
As shown in
Section~\ref{sec::sm} (Fig.~\ref{fig::xsecsm}(b)), the variation of
$\mu_h$ only leads to a very small effect and thus we keep it fixed.

Following the common practice, we write the hadronic cross section
for the individual channels as\footnote{The subscripts $hk$ mark the partons
  in the initial state. Below we sum over all possibilities.}
\begin{eqnarray}
  \sigma_{hk}(s) &=& \sigma_0 C_g^2 \Sigma_{hk}(s)
  \,,
  \label{eq::sigma_hk}
\end{eqnarray}
where $hk\in\{gg,qg,q\bar{q},qq,qq^\prime\}$ ($q$ and $q^\prime$ denote
different light quark flavours) and 
$\sigma_0$ contains the exact dependence on the heavy masses at the
LO. In the SM this is only the top quark mass, while in the MSSM
also the top squark masses $m_{\tilde{t}_1} = m_{\tilde{t}_2} = \msusy$ 
must be taken into account\footnote{The terms of order $M_W^2/m_t^2$ are not
displayed but are accounted for in the numerical evaluation.}:
\begin{eqnarray}
  \sigma_0 &=&\left| \frac{c_\alpha}{s_\beta} \left\{
      \frac{3}{2}\tau_t \left[1 + (1-\tau_t)\right] f(\tau_t)
      - \frac{m_t^2}{2 \msusy^2} 3 \tau_s \left[1 - \tau_s f(\tau_s)\right]
    \right\} + {\cal O}\left(\frac{M_W^2}{m_t^2}\right) \right|^2\,.
\end{eqnarray}
where $\tau_t = 4 m_t^2/M_H^2$, $\tau_s = 4 \msusy^2/M_H^2$ and
\begin{equation}
  f(t) = 
  \left\{\begin{array}{cl}
      \arcsin^2(1/\sqrt{t}), & t \ge 1\,, \\
      - \frac{1}{4} \left(\ln\frac{1 + \sqrt{1-t}}{1 - \sqrt{1-t}} - i\pi \right)^2,
      & t < 0\,. 
    \end{array}
  \right.
\end{equation}

The quantity $\Sigma_{hk}(s)$ in Eq.~(\ref{eq::sigma_hk}), expanded in
$\alpha_s^{(5)}(\mu_s)$, contains the convolutions of the PDFs with
the $n$-loop ($n=0,1,2$) effective-theory partonic cross sections.
In the SM, such effective theory approach provides an excellent
approximation at NNLO~\cite{Harlander:2009mq,Pak:2009dg,Harlander:2010my}.
Once the LO cross section which depends on $\alpha_s^{(5)}(\mu_s)$ is
factored out, we are left with a product of the residual hard coefficient
depending on $\alpha_s^{(5)}(\mu_h)$ and the
remaining part of $\Sigma_{hk}(s)$ which is parameterized in terms of
$\alpha_s^{(5)}(\mu_s)$.
We expand these factors and consistently discard higher orders in
the strong coupling, decomposing the hadronic cross section into
the LO, NLO and NNLO pieces. Introducing
\begin{eqnarray}
  C_g &=& 1 + \frac{\alpha_s^{(5)}(\mu_h)}{\pi} c_g^{(1)}
  + \left(\frac{\alpha_s^{(5)}(\mu_h)}{\pi}\right)^2
  c_g^{(2)}
  + \ldots
  \,,\nonumber\\
  \Sigma &=& \Sigma^{(0)} + \frac{\alpha_s^{(5)}(\mu_s)}{\pi} \Sigma^{(1)}
  + \left(\frac{\alpha_s^{(5)}(\mu_s)}{\pi}\right)^2 \Sigma^{(2)}
  + \ldots  \,,
\end{eqnarray}
we write the production cross section as
\begin{eqnarray}
  \sigma^{\rm NNLO} &=& \sigma_0
  \Bigg\{\Sigma^{(0)} 
    + \frac{\alpha_s^{(5)}(\mu_s)}{\pi}\Sigma^{(1)} 
    + \frac{\alpha_s^{(5)}(\mu_h)}{\pi}2c_g^{(1)}\Sigma^{(0)} 
    + \left(\frac{\alpha_s^{(5)}(\mu_s)}{\pi}\right)^2\Sigma^{(2)} 
    \nonumber\\&&\mbox{}
    + \frac{\alpha_s^{(5)}(\mu_s)}{\pi}\frac{\alpha_s^{(5)}(\mu_h)}{\pi}
    2c_g^{(1)}\Sigma^{(1)}  
    + \left(\frac{\alpha_s^{(5)}(\mu_h)}{\pi}\right)^2
    \left[(c_g^{(1)})^2+2c_g^{(2)}\right]\Sigma^{(0)}   
    + \ldots
    \Bigg\}
    \,,
    \nonumber\\
\end{eqnarray}
where superscripts on $c_g$ or $\Sigma$ denote the loop order of the
corresponding quantity.

As input for the numerical evaluation we use the on-shell top
quark mass $M_t=173.3$~GeV~\cite{:1900yx} and convert it to the 
$\overline{\rm DR}$ scheme using two-loop
SUSY-QCD corrections~\cite{Martin:2005ch}.
By default we use
LO, NLO and NNLO PDFs by MSTW2008 collaboration~\cite{Martin:2009iq}.
The choice of the PDFs determines the values of $\alpha_s^{(5)}(M_Z)$
at the considered order. Using the appropriate $\overline{\rm MS}$ beta
function, we then derive $\alpha_s^{(5)}(\mu_s)$ and $\alpha_s^{(5)}(\mu_h)$.

\begin{figure}[t]
  \centering
  \begin{tabular}{c}
    \includegraphics[width=0.9\linewidth]{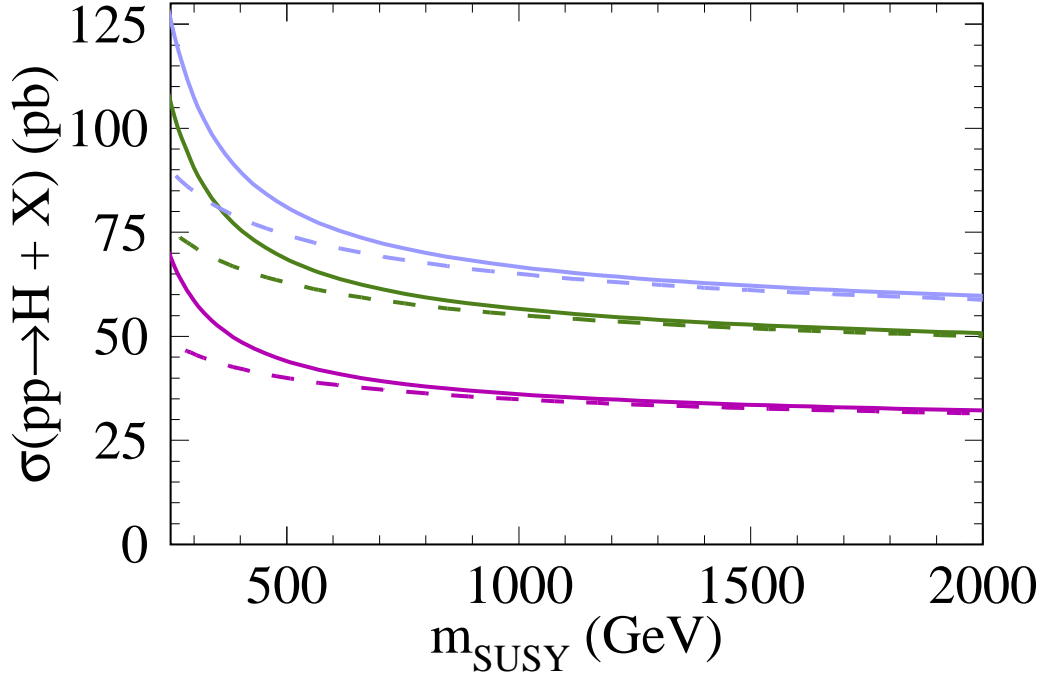}
  \end{tabular}
  \caption[]{\label{fig::degen_xsec}
    LO (lower), NLO (middle)
    and NNLO (upper curves) Higgs production cross section as a function of 
    $m_{\rm susy}$.
    The solid and dashed lines correspond to the MSSM and SM, respectively.
    }
\end{figure}

In Fig.~\ref{fig::degen_xsec} we show the total cross section as a
function of a common supersymmetric mass scale $m_{\rm susy}$ up to
NNLO with $\tan\beta=5$, $\mu_{\rm susy}=200$~GeV,
vanishing mixing angle for the top squarks and a pseudoscalar Higgs boson
mass $M_{A}=1$~TeV. The Higgs boson mass is computed with the help
of the program {\tt H3m}~\cite{Kant:2010tf} as a function of $m_{\rm susy}$
to three-loop accuracy.
In the range of $m_{\rm susy}$ varying between 250~GeV and
2~TeV $M_H$ assumes values between 92~GeV and 117~GeV. Although
experimentally excluded if the SM limits on $M_H$ are assumed, we show the
curves for illustration in the mentioned region for $m_{\rm susy}$.
The mixing angle $\alpha$
is relatively stable so that $\frac{\cos\alpha}{\sin\beta}\approx 0.99$.
The SM results are shown with the dashed lines. They develop 
$m_{\rm susy}$ dependence since we identify the SM Higgs boson mass with
the MSSM value. Note that for $m_{\rm susy} \lsim 1500$~GeV the
corresponding Higgs boson masses are excluded by LEP~\cite{PDG}.
Nevertheless we continue the dashed curves to $m_{\rm susy}=250$~GeV to have a
comparison with the MSSM curves.

One observes that for the large values of
$m_{\rm susy}$ the MSSM curves approach the SM result 
whereas for $m_{\rm susy}\lsim 400-500$~GeV a significant enhancement
of the MSSM cross section is observed.

In order to determine the effect of higher order corrections
it is convenient to define the so-called $K$-factor as the ratio
of the higher order cross section and the Born cross section
with the soft scale fixed at $\mu_s^2=(M_H/2)^2$,
\begin{eqnarray}
  K^{\rm HO} &=& \frac{\sigma^{\rm HO}(\mu_s)}{\sigma^{\rm LO}(\mu_s=M_H/2)}
  \,.
  \label{eq::Kdef}
\end{eqnarray}

The values of the $K$-factor corresponding to Fig.~\ref{fig::degen_xsec}
vary only slightly with $m_{\rm susy}$ and are
$K^{\rm NLO} \approx 1.5$, $K^{\rm NNLO} \approx 1.8$,
very close to their SM counterparts.

We refrain from discussing the renormalization scale dependence since it is
essentially inherited from the SM corrections discussed in
Section~\ref{sec::sm}. 

\begin{figure}[t]
  \centering
  \begin{tabular}{c}
    \includegraphics[width=0.9\linewidth]{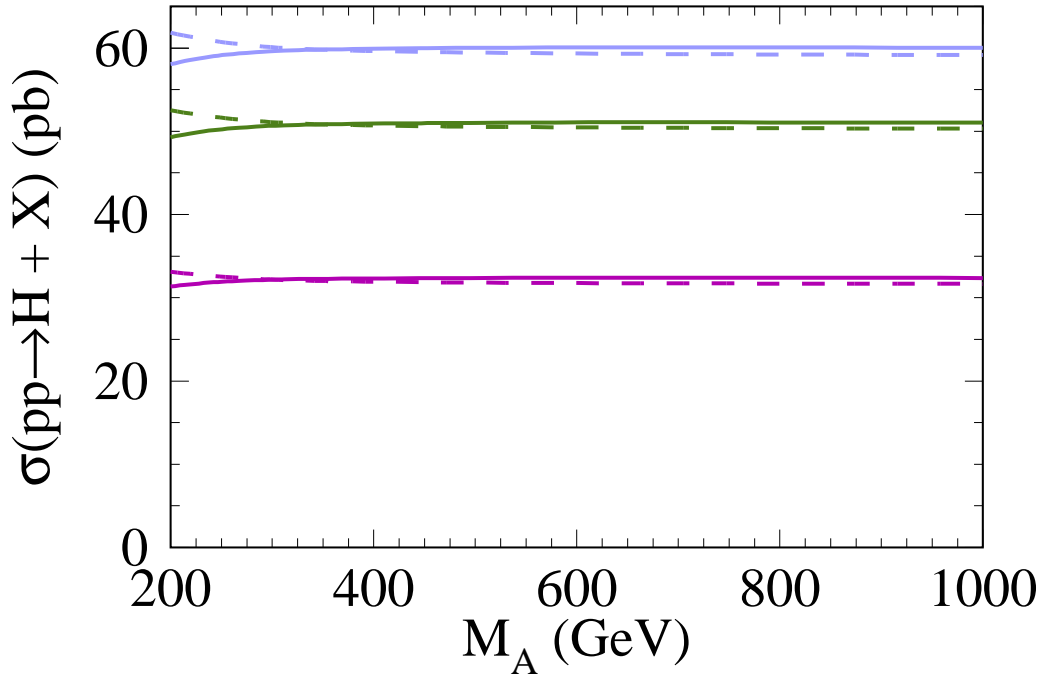}
  \end{tabular}
  \caption[]{\label{fig::nomix_xsec}
    LO (lower), NLO (middle)
    and NNLO (upper curves) Higgs production cross section for the
    ``no-mixing'' scenario as a function of $M_A$.
    The solid and dashed lines correspond to the MSSM and SM, respectively.
    }
\end{figure}

Our following scenario is based on the so-called ``no-mixing'' scheme
as defined in Ref.~\cite{Carena:2002qg}. Keeping
$\tan\beta = 5$ fixed, we vary $M_A$, the mass of the pseudoscalar
Higgs boson, between 200~GeV and 1~TeV, and compute the scalar Higgs
mass and the supersymmetric particle masses with
SOFTSUSY~\cite{Allanach:2001kg}. Averaging the masses of squark, stop and gluino,
we obtain $m_{\text{susy}}\approx 1970$~GeV. In Fig.~\ref{fig::nomix_xsec} we
present the cross section 
computed in this scenario. Here the Higgs boson mass varies
from 114~GeV to 117~GeV, and that  affects the SM cross section
(the dashed lines). Compared to the previous plot, the variation of 
$\alpha$ is more pronounced, reaching $\frac{\cos\alpha}{\sin\beta}\approx 0.95$
at the left boundary, while for the larger values of $M_A$ this ratio
approaches 1. 
The genuine supersymmetric corrections (arising from terms of ${\cal
  O}(1/m_{\rm susy})$ in Eq.~(\ref{eq::C1DR})) are small.
Again, the $K$-factors are very stable in the whole
range of $M_A$ and are $K^{\rm NLO} = 1.6$ and $K^{\rm NNLO} = 1.9$.


\section{\label{sec::concl}Conclusions}

In this letter we have computed supersymmetric NNLO SUSY-QCD corrections
arising from the top quark/squark sector
to the matching coefficient determining the Higgs-gluon coupling in the
effective theory.
Regularization was done in {\tt DRED} with the help of the
so-called $\varepsilon$ scalars which made the matching 
to the five-flavour QCD very delicate.
In particular, we had to renormalize the mass of the $\varepsilon$ scalar and
its coupling to the Higgs boson at one- and two-loop orders, respectively.

The approximation investigated in this paper, assuming zero mixing in
the stop sector, produces significant effects only when $m_{\rm susy}$
is small, $\ll 1$~TeV. However, the effect of zero mixing on
Higgs mass together with the current experimental bounds on this mass
exclude this possibility. We expect that a more general result,
allowing for a large mixing angle, will not be so severely constrained,
and enhancement of the cross section by as much as 50\% (as in the 
left side of  Fig.~\ref{fig::degen_xsec}) may be allowed.

Analyzing the higher order corrections, in the MSSM we observe
the same pattern as in the SM: large NLO corrections which further
increase at NNLO.
For $m_{\rm susy}\gsim 1$~TeV the genuine supersymmetric
corrections are suppressed and the difference from the SM is mainly due to the
factor $\frac{\cos\alpha}{\sin\beta}$ in Eq.~(\ref{eq::C1DR}).
However, the MSSM $K$-factors are very similar to their SM counterparts.
This confirms the estimate of Ref.~\cite{Harlander:2003kf} based on
the complete NLO corrections to the matching coefficient and
the NNLO calculation in the effective theory.

Our calculation has been performed in the limit of degenerate supersymmetric
masses. In order to explore all interesting regions of the supersymmetric
parameter space it would be necessary to relax this restriction. Since a
three-loop calculation with many different mass scales is currently not
feasible it is more promising to employ a strategy based on asymptotic
expansions which has been applied successfully to predict the lightest
MSSM Higgs boson mass in Ref.~\cite{Kant:2010tf}.



\section*{Acknowledgements}

This work was supported by the DFG through the SFB/TR~9 ``Computational
Particle Physics'' and the Graduiertenkolleg ``Hochenergiephysik und
Teilchenastrophysik''.




\end{document}